\documentclass[5p,times,twocolumn]{elsarticle}
\usepackage{amsmath,amssymb}
\biboptions{comma,sort&compress}
\usepackage[utf8]{inputenc}
\usepackage{graphicx}
\usepackage{bm}
\usepackage{hyperref} 
\hypersetup{colorlinks = true,linkcolor = blue,anchorcolor =red,citecolor = blue,filecolor = red,urlcolor = red,pdfauthor=author}
\usepackage[dvipsnames]{xcolor}

\journal{Physics Letters B}

\begin{document}

\begin{frontmatter}

\title{Metric anisotropies and nonequilibrium attractor for expanding plasma}

\author[niser]{Nisarg Vyas}
\ead{nisarg.vyas@niser.ac.in}
\author[tifr]{Sunil Jaiswal\corref{cor1}}
\ead{sunil.jaiswal@tifr.res.in}
\author[niser]{Amaresh Jaiswal\corref{cor1}}
\ead{a.jaiswal@niser.ac.in}

\cortext[cor1]{Corresponding author}

\address[niser]{School of Physical Sciences, National Institute of Science Education and Research, An OCC of Homi Bhabha National Institute, Jatni 752050, India}
\address[tifr]{Department of Nuclear and Atomic Physics, Tata Institute of Fundamental Research, Mumbai 400005, India}

\date{\today}

\begin{abstract} 
We consider the evolution of a system of chargeless and massless particles in an anisotropic space-time given by the Bianchi type I metric. Specializing to the axis-symmetric case, we derive the framework of anisotropic hydrodynamics from the Boltzmann equation in the relaxation-time approximation. We consider the case of the axis-symmetric Kasner metric and study the approach to the emergent attractor in near and far-off-equilibrium regimes. Further, by relaxing the Kasner conditions on metric coefficients, we study the effect of expansion geometries on the far-off-equilibrium attractor and discuss its implications in the context of relativistic heavy-ion collisions.
\end{abstract}

\begin{keyword}
Relativistic kinetic theory\sep Hydrodynamic models \sep Anisotropic spacetime \sep Relativistic heavy-ion collisions
\end{keyword}

\end{frontmatter}


\section{Introduction}

In the last two decades, there have been major advances in the formulations of relativistic hydrodynamics and its application in relativistic heavy-ion collisions \cite{Muronga:2003ta, York:2008rr, Denicol:2010xn, Heinz:2013th, Jaiswal:2014isa, Braun-Munzinger:2015hba, Jeon:2015dfa, Chattopadhyay:2014lya, Jaiswal:2016hex, Florkowski:2017olj}. It has been well established that the strongly interacting hot and dense matter, created in high energy heavy-ion collisions, is highly anisotropic in momentum space at very early times resulting in large pressure anisotropies at the center of the fireball \cite{Martinez:2008di, Florkowski:2009sw}. Moreover, this anisotropy increases towards the peripheral regions of the fireball which are at relatively lower temperatures \cite{Song:2009gc, Martinez:2012tu}. Traditional viscous hydrodynamic formulation, which assumes small deviations from equilibrium, is incapable of capturing such large momentum space anisotropy and leads to negative longitudinal pressure if one considers early times and/or colder regions of the plasma \cite{Torrieri:2008ip, Martinez:2009mf, Habich:2014tpa, Bhatt:2011kx, Bhatt:2011kr}. In order to overcome these issues, the framework of anisotropic hydrodynamics (aHydro) was developed \cite{Martinez:2010sc, Florkowski:2010cf, Martinez:2010sd, Ryblewski:2011aq, Nopoush:2015yga, Ryblewski:2012rr, Bazow:2013ifa, Tinti:2013vba, Nopoush:2014pfa, Molnar:2016vvu}; see Ref.~\cite{Strickland:2014pga} for a comprehensive review. The framework of aHydro has recently been applied quite successfully to explain the experimental observables in relativistic heavy-ion collisions \cite{Alqahtani:2017jwl, Alqahtani:2017tnq, Alqahtani:2020paa}.

In an earlier work by one of the authors, an alternate derivation of aHydro equations was proposed by considering non-interacting static fluid in an anisotropic space-time, given by Bianchi type I metric \cite{Dash:2017uem}. Expressions for components of the energy-momentum tensor and conserved currents, such as number density, energy density, and pressure components were calculated and shown to be identical to that obtained within aHydro framework when one considers axis-symmetric Bianchi type I metric. The formulation considered free-streaming particles where the momentum anisotropy was completely determined by the anisotropy in the metric. Microscopic interactions among the medium constituents are expected to result in the evolution of the momentum anisotropy such that the system approaches an equilibrium state which is isotropic in momentum space. The inclusion of microscopic interactions in this framework is therefore important to understand the out-of-equilibrium evolution of momentum anisotropy and the subsequent approach to equilibrium. 

In this article, we present an alternative derivation of aHydro equations by considering interacting static fluid in a locally anisotropic background space-time metric, given by Bianchi type I metric. We further consider the special case of axis-symmetric Kasner metric and discuss the implications of our results in the context of aHydro and the emergent nonequilibrium attractor%
    \footnote{We shall refer to the solution which originates from the ``free-streaming stable fixed point'' as ``nonequilibrium'', ``early-time'', or ``far-off-equilibrium'' attractor \cite{Blaizot:2019scw, Jaiswal:2022udf}.}. 
The study of the early-time attractor is essential in understanding the emergent universal behavior in far-off-equilibrium regime \cite{Heller:2013fn,  Heller:2015dha, Kurkela:2015qoa, Blaizot:2017lht, Romatschke:2017vte, Spalinski:2017mel,  Strickland:2017kux, Romatschke:2017acs, Behtash:2017wqg, Blaizot:2017ucy, Romatschke:2017ejr, Kurkela:2018wud, Mazeliauskas:2018yef, Strickland:2018ayk,  Jaiswal:2019cju, Kurkela:2019set,  Behtash:2019txb, Heinz:2019dbd, Blaizot:2019scw, Blaizot:2020gql, Behtash:2020vqk, Blaizot:2021cdv, Ambrus:2021sjg, Chattopadhyay:2021ive, Jaiswal:2021uvv, Jaiswal:2022udf}. The framework presented in this work allows for different expansion geometries, and thus provides a suitable platform to study the existence of far-off-equilibrium attractor. By relaxing the Kasner conditions on metric parameters, we analyze the effect of expansion geometries on the far-off-equilibrium attractor. We find that the far-off-equilibrium attractor is governed completely by the relative expansion rates along the longitudinal and transverse axes. We also discuss the relevance of these findings in the context of relativistic heavy-ion collisions.

\section{The metric}

The line element for spatially homogeneous and anisotropic Bianchi type-I space-time is given by \cite{Misner:1967uu,Jacobs:1968zz,Thorne:1967zz}
\begin{equation}\label{eq:LineElement}
ds^2=dt^2 - A^2(t)\,dx^2 - B^2(t)\,dy^2 - C^2(t)\,dz^2,
\end{equation}
where $A(t)$, $B(t)$, and $C(t)$ are scale factors for expansion along $x$, $y$, and $z$ directions, respectively. The metric is diagonal and can be written as
\begin{equation}\label{eq:metric}
g_{\mu\nu}={\rm diag}\left[1,-A^2(t),-B^2(t),-C^2(t)\right],
\end{equation}
and its inverse is given by $g^{\mu\nu}$. The determinant of the metric will be useful in defining the invariant integral measure: $\sqrt{-g}\equiv\sqrt{-{\rm det}(g_{\mu\nu})} = ABC$. The above metric has only six non-zero Christoffel symbols
\begin{equation}\label{eq:Christoffel}
\Gamma_{xt}^{x} = \Gamma_{tx}^{x} = \frac{\dot A}{A}, \quad \Gamma_{yt}^{y} = \Gamma_{ty}^{y} = \frac{\dot B}{B}, \quad \Gamma_{tz}^{z} = \Gamma_{zt}^{z} = \frac{\dot C}{C},
\end{equation}
where $\dot{A}\equiv dA/dt$. The scale factors $A$, $B$, and $C$ are time dependent quantities unless specified otherwise. In this article, we will restrict ourselves to the axis-symmetric case where $A=B$. Later, we shall also consider the situation when Bianchi type I metric is a solution of the vacuum Einstein equation known as the Kasner metric.

\vspace{-.2cm}
\section{Spheroidal energy-momentum tensor}
\vspace{-.1cm}

We specify a general tensor basis \cite{Martinez:2012tu, Strickland:2014pga, Alqahtani:2017mhy},
\begin{align}
X^\mu_0 &\equiv u^\mu = \left[ 1,0,0,0 \right], \label{umu_LRF}\\
X^\mu_1 &\equiv x^\mu = \left[ 0,1/A(t),0,0 \right], \label{xmu_LRF}\\
X^\mu_2 &\equiv y^\mu = \left[ 0,0, 1/A(t),0 \right], \label{ymu_LRF}\\
X^\mu_3 &\equiv z^\mu = \left[ 0,0,0,1/C(t) \right]. \label{zmu_LRF}
\end{align}
Note that the four-vectors in the above equations are orthogonal in all frames. The unit time-like four-vector $X^\mu_0$, specified in Eq.~\eqref{umu_LRF}, may be associated with the four-velocity of the fluid, which we assume to be globally static. Note that the basis vectors, defined in Eqs.~\eqref{umu_LRF}-\eqref{zmu_LRF}, scale as metric expands or contracts in the respective directions and differ from similar definitions in Refs.~\cite{Martinez:2012tu, Strickland:2014pga, Alqahtani:2017mhy} in this respect. 

Using the basis vectors defined in Eqs.~\eqref{umu_LRF}-\eqref{zmu_LRF}, one can form a tensor of any rank. For instance, it can be checked that the metric tensor itself can be written as
\begin{equation}\label{metric_X}
g^{\mu\nu} = u^\mu u^\nu - \sum_{i=1}^3 X_i^\mu X_i^\nu.
\end{equation}
Energy-momentum tensor is a symmetric two rank tensor and can be expressed in terms of the basis vectors as
\begin{equation}\label{Tmunu_basis}
T^{\mu\nu} = t_{00}\,g^{\mu\nu} + \sum^3_{i=1} t_{ii}X^{\mu}_i X^{\nu}_i + \sum^3_{\substack{\alpha,\,\beta=0 \\ \alpha>\beta}} t_{\alpha\beta}\left( X^{\mu}_\alpha X^{\nu}_\beta + X^{\nu}_\alpha X^{\mu}_\beta \right).
\end{equation}
Next, we specialize to the case of spheroidal anisotropic hydrodynamics where $P_x=P_y\equiv P_T$ and $P_z\equiv P_L$, with $P_x$, $P_y$ and $P_z$ being the pressure along $x$, $y$ and $z$ directions, respectively. In this case, solving for the coefficients $t_{\alpha\beta}$, we obtain \cite{Martinez:2012tu, Strickland:2014pga, Alqahtani:2017mhy}
\begin{equation}\label{Tmunu_eq}
T^{\mu\nu} = (\epsilon+P_T)u^{\mu}u^{\nu} - P_T g^{\mu\nu} + (P_L -P_T)z^{\mu}z^{\nu}.
\end{equation}
It is important to note that the pressure anisotropy can have contributions from anisotropic momentum distribution.

For a system of massless particles, it was shown in Ref.~\cite{Dash:2017uem} that a collisionless evolution of an initially equilibrated system in anisotropically expanding space-time metric exhibits a distribution identical to Romatschke-Strickland form of the momentum distribution function,
\begin{equation}\label{ani_dist}
f_{\rm RS}(x,p) \equiv f_{\rm eq}\left( \frac{\sqrt{{\bf p}^2 + \xi(x)\, p_z^2}}{\Lambda(x)} \right),
\end{equation}
where $f_{\rm eq}$ represents the equilibrium distribution function. In this case, an asymptotic observer attributes the anisotropy in the momentum space distribution to anisotropy introduced by the metric. The quantity $\xi(x)$, appearing in Eq.~\eqref{ani_dist}, is the anisotropy parameter which leads to prolate, isotropic, and oblate distributions for $-1<\xi<0$, $\xi=0$ and $0<\xi<\infty$, respectively
    \footnote{The relation between anisotropy parameter $\xi$ and anisotropy in momentum space is: $$\xi = \frac{C^2(t)/C^2(t_0)}{A^2(t)/A^2(t_0)}-1.$$}. 
Furthermore, $\Lambda(x)$ is the local scale parameter which can be interpreted as the temperature in the isotropic limit, i.e., for $\xi\to0$. We use the notation ${\bf p}^2 \equiv p_x^2+p_y^2+p_z^2$ with $p_x$, $p_y$ and $p_z$ being the particle momenta along $x$, $y$ and $z$ directions, respectively. We consider the equilibrium distribution function, $f_{\rm eq}$, to have the form
\begin{equation}\label{eq_dist}
f_{\rm eq} (q) = \left[\exp(q) + \lambda\right]^{-1},
\end{equation}
where $\lambda=0,~1,~-1$ for Maxwell-Boltzmann, Fermi-Dirac or Bose-Einstein distributions, respectively. In the present work, we consider the evolution of momentum-space anisotropy, due to microscopic interactions, in the presence of anisotropic background metric. Here we assume the Romatschke-Strickland ansatz for the distribution function, motivated by the analysis of Ref.~\cite{Dash:2017uem}.

The energy-momentum tensor can be expressed in terms of the distribution function as 
\begin{equation}\label{eq:SET}
T^{\mu\nu} = \int dP \,p^\mu\,p^\nu\,f\left(x,p\right),
\end{equation}
where the invariant momentum integral measure is defined as $dP\equiv \sqrt{-g}\, d^3 \textbf{p}/\left[ (2\pi)^3 p^0\right]$. Using the ansatz~\eqref{ani_dist} for the anisotropic distribution function, the components of energy-momentum tensor can be obtained as \cite{Martinez:2009ry}
\begin{align}
\epsilon &= \mathcal{R}(\xi)\, \epsilon_{\rm eq}(\Lambda), \qquad \,
\mathcal{R}(\xi)=\frac{1}{2} \left(\frac{1}{1+\xi}+\frac{\tan^{-1}\!(\sqrt{\xi})}{\sqrt{\xi}}\right),
\label{eq:aniso_e}\\
P_{L} &= \mathcal{R}_{L}(\xi)\, P_{\rm eq}(\Lambda), \quad
\mathcal{R}_{L}(\xi) =\frac{3}{\xi}\left(\mathcal{R}(\xi) -\frac{1}{\xi+1}\right),
\label{eq:aniso_PL}\\
P_{T} &= \mathcal{R}_{T}(\xi)\, P_{\rm eq}(\Lambda), \quad
\mathcal{R}_{T}(\xi) =\frac{3}{2 \xi}\left(\frac{1+\left(\xi^{2}-1\right) \mathcal{R}(\xi)}{\xi+1} \right), \label{eq:aniso_PT}
\end{align}
where $\epsilon_{\rm eq}(\Lambda)$ and $P_{\rm eq}(\Lambda)$ is the equilibrium energy density and pressure, respectively, obeying the relation $\epsilon_{\rm eq}(\Lambda) = 3P_{\rm eq}(\Lambda)$. It can be shown that similar result also holds for spheroidal net number density, $n=n_{\rm eq}(\Lambda)/\sqrt{1+\xi}$. The formulation of anisotropic hydrodynamics now amounts to finding the evolution of the parameters $\xi$ and $\Lambda$, which can be achieved by taking appropriate moments of the Boltzmann equation.

\vspace{-.3cm}
\section{Boltzmann equation and its moments}
\vspace{-.2cm}

In this section, we consider the relativistic Boltzmann equation in general space-time background metric, which can be expressed as
\begin{equation}\label{eq:Boltz3}
p^{\mu}\partial_{\mu}f-\Gamma^{\mu}_{\alpha\beta}p^{\alpha}p^{\beta}\frac{\partial f}{\partial p^{\mu}}=C[f],
\end{equation}
where $\partial_\mu\equiv \partial/\partial x^\mu$ and $C[f]$ is the collision kernel. In the following, we consider relaxation-time approximation for the collision term,
\begin{equation}\label{RTA}
C[f]=- \frac{u \cdot p }{\tau_{R}}\,(f - f_{\rm eq}),
\end{equation}
which assumes that when perturbed slightly out of equilibrium, the system approaches equilibrium exponentially with a timescale $\tau_R$. In the case of axis-symmetric expansion with symmetry in the transverse direction, i.e., $A(t) = B(t)$, Eq.~\eqref{eq:Boltz3} in the relaxation-time approximation takes the form,
\begin{equation} \label{eq:Boltz4}
p^{\mu}\partial_{\mu}f -  2 p^{0} \left[\frac{\dot{A}}{A}\left(p^{x}\frac{\partial f}{\partial p^{x}} + p^{y}\frac{\partial f}{\partial p^{y}}\right) + \frac{\dot{C}}{C} p^{z}\frac{\partial f}{\partial p^{z}} \right] = - \frac{u \cdot p }{\tau_{R}}(f - f_{\rm eq}).
\end{equation}
The evolution of parameters $\Lambda$ and $\xi$ can be obtained by assuming that the anisotropic distribution function, Eq.~\eqref{ani_dist}, satisfies the above equation. The evolution equations for $\Lambda$ and $\xi$ are then derived by considering suitable moments of Eq.~\eqref{eq:Boltz4}.

Energy-momentum conservation requires that we have $D_\mu T^{\mu\nu}\equiv\partial_\mu T^{\mu\nu} + \Gamma^\mu_{\mu\sigma}T^{\sigma\nu} + \Gamma^\nu_{\mu\sigma}T^{\mu\sigma}=0$, where $D_\mu$ is the covariant derivative. We find that the first moment of the Boltzmann equation, defined as considering $\int dP\,p^\nu$ on both sides of Eq.~\eqref{eq:Boltz4}, leads to energy-momentum conservation \cite{cercig_Kremer}, provided the collision kernel in the relaxation-time approximation satisfies 
\begin{equation}
- \int dP \,p^\nu \frac{u \cdot p }{\tau_{R}}\,(f - f_{\rm eq}) = 0.
\end{equation}
For the spatial components ($\nu=1,2,3$), this can be easily verified by noticing that in the local fluid rest frame, the integral vanishes because it is an odd function in $p^x$, $p^y$, or $p^z$. In the above equation, by choosing the Landau frame condition ($u_\alpha T^{\alpha\beta}=\epsilon u^\beta$) to define the fluid velocity, the temporal component ($\nu=0$) leads to the so-called Landau matching condition,
\begin{equation}\label{Landau_match}
\epsilon(\xi, \Lambda) = \epsilon_{\rm eq} (T).
\end{equation}
For conformal case with $\epsilon_{\rm eq} (T)\sim T^4$, the above equation, along with Eq.~\eqref{eq:aniso_e}, leads to 
\begin{equation}\label{eq:T_and_Lambda}
T = [\mathcal{R}(\xi)]^{1/4}\,\Lambda.
\end{equation}
The temperature determined from the above equation is referred to as the effective temperature. Considering the projection along $u^\mu$ of the first moment of the Boltzmann equation~\eqref{eq:Boltz4}, we get
\begin{equation}\label{first_moment}
\frac{\mathcal{R}^{\prime}}{\mathcal{R}} \frac{\partial \xi}{\partial t} + \frac{4}{\Lambda} \frac{\partial \Lambda}{\partial t} = -\left(2 \frac{\dot A}{A} + \frac{\dot C}{C}\right) \left[1+\frac{1}{3} \frac{\mathcal{R}_L}{\mathcal{R}} \right],
\end{equation}
where $\mathcal{R}^{\prime}\equiv\partial\mathcal{R}/\partial\xi$.

In order to close the set of equations for $\Lambda$ and $\xi$, we need one more evolution equation. While there are various choices of moments to close this set of equations \cite{Molnar:2016gwq}, the conclusions of the present article are insensitive to this moment choice. In this work, we consider the second moment of the Boltzmann equation as it captures the correct near-equilibrium limit if one uses the conformal expression for the relaxation time, $\tau_R=5\bar{\eta}/T$ \cite{Alqahtani:2017mhy}. Here $\bar{\eta}\equiv\eta/s$ is the ratio of shear viscosity to entropy density, also known as specific shear viscosity. Similar to the first moment, the second moment is defined by considering $\int dP\,p^\alpha p^\beta$ on both sides of the Boltzmann equation~\eqref{eq:Boltz4}. We first consider the $zz$ projection of the second moment, defined as $z_\alpha z_\beta \int dP\,p^\alpha p^\beta (\cdots)$, where $z^\mu$ in the local rest frame is given in Eq.~\eqref{zmu_LRF}. The $zz$ projection of the second moment leads to
\begin{equation}\label{proj_zz}
\frac{\mathcal{S}_L^\prime}{\mathcal{S}_L} \frac{\partial \xi}{\partial t}+\frac{5}{\Lambda} \frac{\partial \Lambda}{\partial t} + \left( 2\frac{\dot A}{A} + 3\frac{\dot C}{C} \right) =\frac{1}{\tau_{R}} \left[\frac{\mathcal{R}^{5 / 4}}{\mathcal{S}_{L}}-1\right].
\end{equation}
On the other hand, both $xx$ and $yy$ projection, defined in a similar way using Eqs.~\eqref{xmu_LRF} and \eqref{ymu_LRF}, respectively, gives
\begin{equation}\label{proj_xx-yy}
\frac{\mathcal{S}_T^\prime}{\mathcal{S}_T} \frac{\partial \xi}{\partial t} +\frac{5}{\Lambda} \frac{\partial \Lambda}{\partial t} + \left( \frac{4\dot A}{A} + \frac{\dot C}{C} \right) =\frac{1}{\tau_{R}} \left[\frac{\mathcal{R}^{5 / 4}}{\mathcal{S}_{T}}-1\right].
\end{equation}
In the above equations, the quantities $\mathcal{S}_L$ and $\mathcal{S}_T$ are defined as
\begin{equation}\label{SLST}
\mathcal{S}_L (\xi) \equiv \frac{1}{(1+\xi)^{3 / 2}}, \quad \mathcal{S}_T (\xi) \equiv \frac{1}{\sqrt{1+\xi}}.
\end{equation}
A linear combination of Eqs.~\eqref{proj_zz} and \eqref{proj_xx-yy} leads to
\begin{equation}\label{second_moment}
\frac{1}{1+\xi} \frac{\partial \xi}{\partial t} + 2\left(\frac{\dot A}{A} - \frac{\dot C}{C} \right) + \frac{\mathcal{R}^{5/4}}{\tau_R} \xi \sqrt{1+\xi}=0.
\end{equation}
The equation of motion of anisotropic hydrodynamics is now completely determined by the simultaneous solutions of Eqs.~\eqref{first_moment} and \eqref{second_moment}. The hydrodynamic quantities are then determined using Eqs.~\eqref{eq:aniso_e}, \eqref{eq:aniso_PL}, \eqref{eq:aniso_PT} and \eqref{eq:T_and_Lambda}.

\vspace{-.2cm}
\section{The Kasner metric}
\vspace{-.2cm}

We restrict ourselves to a subclass of Bianchi type-I metric which are classical vacuum solutions of the Einstein equation: the Kasner metric. In this case, the metric scale factors have the form
\begin{equation}\label{Kasner}
A = t^a, \qquad B = t^b, \qquad C = t^c,
\end{equation}
where the parameters $(a,b,c)$ are restricted by the conditions
\begin{equation}\label{Kasner_cond}
a + b + c = 1, \qquad a^2 + b^2 + c^2 = 1.
\end{equation}
From Eq.~\eqref{Kasner}, we find that
\begin{equation}
\frac{\dot A}{A} = \frac{a}{t}, \qquad \frac{\dot B}{B} = \frac{b}{t}, \qquad \frac{\dot C}{C} = \frac{c}{t},
\end{equation}
such that
\begin{equation}
\frac{\dot A}{A} + \frac{\dot B}{B} + \frac{\dot C}{C} = \frac{1}{t}.
\end{equation}
We notice that owing to the above condition, the first moment of the Boltzmann equation which corresponds to energy-momentum conservation, Eq.~\eqref{first_moment}, does not depend on metric parameters. Kasner space-time has also been considered in Ref.~\cite{Pruseth:2020gwf, Pruseth:2022xsd} in the context of relativistic hydrodynamics.

Restricting further to the azimuthally symmetric case, i.e., $a=b$, there are only two possibilities for the Kasner metric for $(a,b,c)$,
\begin{equation}\label{cases}
{\rm Case~I:~}  \left( 0,0,1 \right) , \qquad
{\rm Case~II:~} \left(\frac{2}{3},\frac{2}{3},-\frac{1}{3}\right).
\end{equation}
We note that the first case is analogous to the Milne metric which is the natural choice of coordinate system in case of longitudinally expanding, boost-invariant, Bjorken flow. This case has been studied extensively in the context of anisotropic hydrodynamics. In addition, we find that the second case represents transverse expansion and longitudinal contraction. We study the attractor behavior in both cases and find universal features which are common in these scenarios.

\subsection{Evolution of anisotropy parameter \texorpdfstring{$\xi$}{}}

For the two azimuthally symmetric Kasner cases, the first moment of the Boltzmann equation corresponding to energy-momentum conservation, Eq.~\eqref{first_moment}, reduces to
\begin{equation}\label{first_moment_Kasner}
\frac{\mathcal{R}^{\prime}}{\mathcal{R}} \frac{\partial \xi}{\partial t} + \frac{4}{\Lambda} \frac{\partial \Lambda}{\partial t}  = -\frac{1}{t}\left[1+\frac{1}{3} \frac{\mathcal{R}_L}{\mathcal{R}} \right].
\end{equation}
On the other hand, for these two cases, we find that the second moment, Eq.~\eqref{second_moment}, reduces to
\begin{equation}\label{second_moment_Kasner}
\frac{1}{1+\xi} \frac{\partial \xi}{\partial t} + \chi\frac{2}{t} + \frac{\mathcal{R}^{5/4}}{\tau_R} \xi \sqrt{1+\xi}=0,
\end{equation}
where $\chi\equiv a-c$, and $\chi=-1,\,+1$ for the first and second Kasner case in Eq.~\eqref{cases}, respectively. We note that Eqs.~\eqref{first_moment_Kasner} and \eqref{second_moment_Kasner} completely determine the evolution in axis-symmetric Kasner background within the framework of anisotropic hydrodynamics. It is important to observe that in the Kasner case I, these equations reduce to those obtained for Bjorken expanding system in Milne coordinates \cite{Martinez:2010sc, Strickland:2017kux}.

We shall now recast the above equation in terms of rescaled time variable: $\bar\tau\equiv t/\tau_R$. Using Eq.~\eqref{eq:T_and_Lambda} for the effective temperature, we obtain,
\begin{align}\label{temp_deriv}
\frac{1}{T} \frac{\partial T}{\partial t} &=\frac{1}{\Lambda} \frac{\partial \Lambda}{\partial t}+\frac{1}{4} \frac{R^{\prime}}{R} \frac{\partial \xi}{\partial t}
= -\frac{1}{4t}\left(1+\frac{1}{3} \frac{\mathcal{R}_L}{\mathcal{R}} \right),
\end{align}
where we have used Eq.~\eqref{first_moment_Kasner} for the last equality.

For conformal system, the relaxation time is given by $\tau_R=5\bar{\eta}/T$. Therefore the derivative, 
\begin{equation}\label{chain_rule}
\frac{\partial}{\partial t} = \frac{\partial \bar{\tau}}{\partial t} \frac{\partial}{\partial \bar{\tau}} = \frac{1}{\tau_{R}}\left(1+\frac{t}{T} \frac{\partial T}{\partial t}\right) \frac{\partial}{\partial \bar{\tau}}
=\frac{1}{\tau_{R}}\!\left(\frac{9\mathcal{R}-\mathcal{R}_L}{12\mathcal{R}}\right)\! \frac{\partial}{\partial \bar{\tau}}.
\end{equation}
Equation \eqref{second_moment_Kasner} using the above equation simplifies to,
\begin{equation}\label{eq:attractor}
\left(\frac{9\mathcal{R}-\mathcal{R}_L}{12(1+\xi)\mathcal{R}} \right) \frac{\partial \xi}{\partial \bar{\tau}} + \chi \frac{2}{\bar{\tau}}  +  \mathcal{R}^{5/4}\xi\sqrt{1+\xi} = 0.
\end{equation}
The above equation represents the evolution of $\xi$ as a function of $\bar{\tau}$ and, unlike Eq.~\eqref{second_moment_Kasner}, it does not depend on $\Lambda$ evolution via $\tau_R$. As a consequence, $\xi(\bar\tau)$ is independent of specific shear viscosity and initial temperature\footnote{Any choice of moment closure leads to such decoupling for the conformal case.}.

\vspace{-.2cm}
\subsection{Gradient expansion}

\begin{figure}
    \centering
    \includegraphics[width=.74\linewidth]{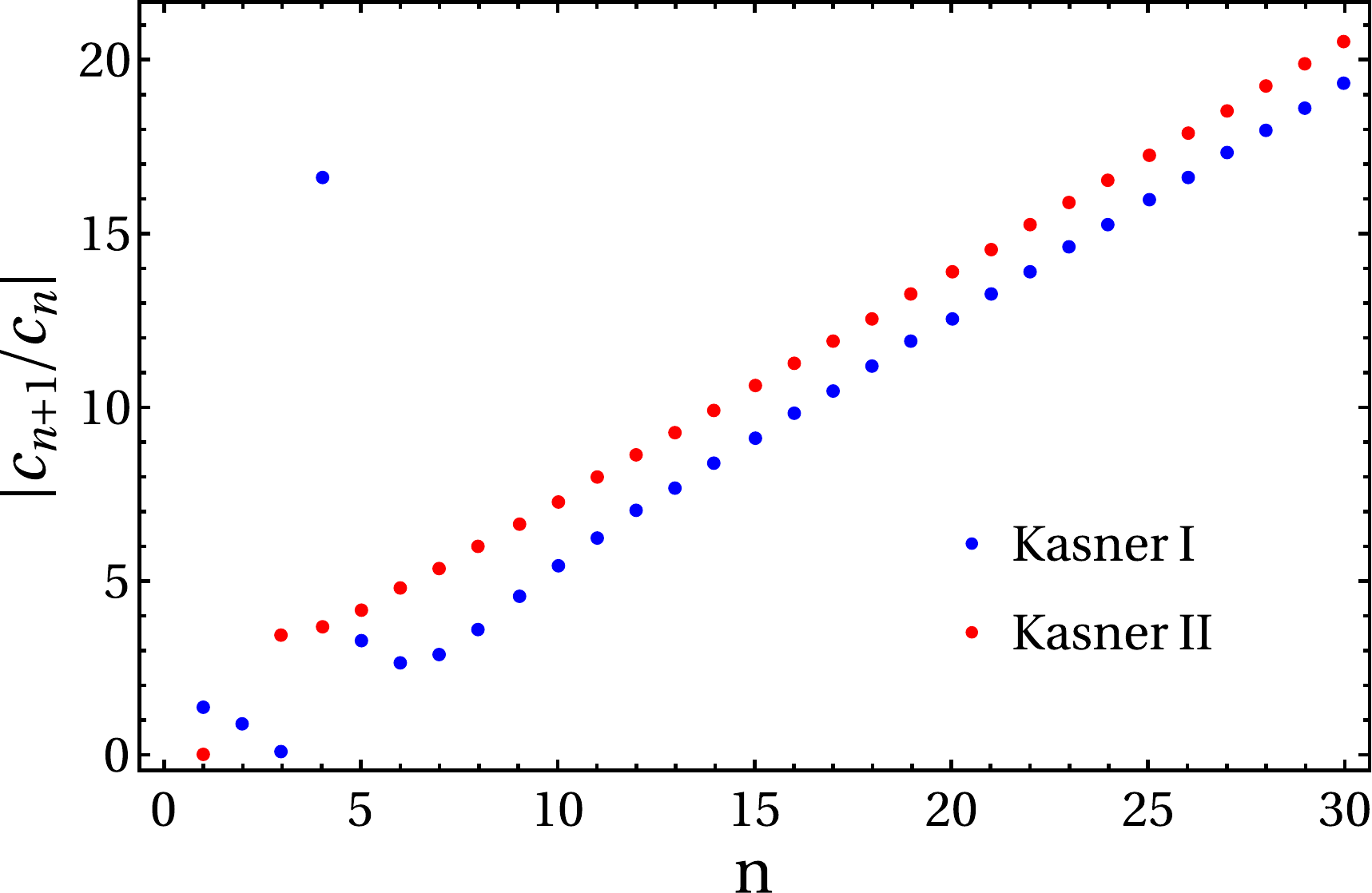}
    \vspace{-.2cm}
    \caption{Factorial behavior of coefficients in Eq.~\eqref{series}  for two Kasner cases.}%
    \label{fig:GE}%
      \vspace{-.2cm}
\end{figure}

In this section, we study the near-equilibrium regime of the above equations by doing a gradient expansion around equilibrium. We consider the solution of $\xi$ corresponding to large $\bar{\tau}$ expansion. This is equivalent to assuming series expansion of $\xi$ in powers of $1/\bar{\tau}$,
\begin{equation}\label{series}
\xi(\bar{\tau})=\sum_{n=0}^\infty \frac{c_n}{\bar{\tau}^n}.
\end{equation}
Equation~\eqref{eq:attractor} can be written as
\begin{align}\label{eq:attractor_GE}
\left[\!\frac{\xi(1\!+\xi)(9\mathcal{R}-\mathcal{R}_L)}{12} \!\right]\! \frac{\partial \xi}{\partial \bar{\tau}} &+ \chi \frac{2\xi(1\!+\xi)^2 \mathcal{R}}{\bar{\tau}} 
+ \xi^2(1\!+\xi)^{5/2}\mathcal{R}^{9/4} \!= 0.
\end{align}
Substituting the ansatz~\eqref{series} in above equation and equating terms with same powers of $1/\bar{\tau}$, we readily obtain the first two coefficients,
\begin{equation}
c_0 = 0, \quad c_1 = -2\chi.
\end{equation}
Indeed this is expected because the anisotropy $\xi$ in Eq.~\eqref{ani_dist} can be attributed to dissipative effects in the large $\bar{\tau}$ limit and gets contribution starting at first order in gradients. It should be noted that the equations of Kasner I case are identical to aHydro equations undergoing Bjorken expansion, and thus the gradient series generated is identical to one found in \cite{Florkowski:2016zsi}. In Fig.~\ref{fig:GE}, we plot the ratio $|c_{n+1}/c_n|$ of the consecutive coefficients in Eq.~\eqref{series}, for both Kasner cases. The linear growth of this ratio, as a function of $n$, demonstrates factorial growth of coefficients $c_n$ for both cases, indicating the divergence of the  gradient series~\eqref{series}. 

The relativistic Navier-Stokes solution is the near-equilibrium behavior of all causal hydrodynamic theories \cite{Heller:2013fn, Heller:2015dha}. It has been shown in earlier works that perturbations in Israel-Stewart hydrodynamic theories decay exponentially, if perturbed around the Navier-Stokes solution \cite{Behtash:2019txb, Jaiswal:2019cju}. To check if this holds for the ahydro equations, we consider perturbation in Eq.~\eqref{eq:attractor} about the linear term of the gradient expansion, i.e., $\xi=c_1/\bar{\tau}+\delta\xi$. Keeping terms linear in $\delta\xi$ and up to first-order in gradients, we obtain
\begin{equation}\label{liner_pert}
\frac{d}{d\bar{\tau}}\delta\xi-\left(\frac{1+\chi}{\chi}\right)\frac{\delta\xi}{\bar{\tau}}+\frac{3}{2}\delta\xi=0.
\end{equation}
The solution to the above equation is readily obtained as
\begin{equation}\label{NS_decay}
\delta\xi\sim e^{-(3/2)\bar{\tau}}\,\bar{\tau}^{(1+\chi)/\chi},
\end{equation}
from which we conclude that the Lyapunov exponent for both Kasner cases is $-3/2$. Also, note that the exponent of $\bar\tau$ is $(1+\chi)/\chi = 0, 2$ for Kasner I and Kasner II, respectively.

\subsection{Fixed points and far-off-equilibrium attractor}

The free-streaming fixed points can be easily obtained from Eq.~\eqref{second_moment_Kasner}. In the free streaming limit, $\tau_R \to \infty$ and the solution of the resulting free-streaming equation is simply
\begin{equation}
    \xi_{\rm FS}(t) = -1 + \left(\frac{t_{\rm in}}{t}\right)^{{2\chi}} (1+\xi_{\rm in}),
\end{equation}
where $\xi_{\rm in}$ is the anisotropy at the initial time $t_{\rm in}$. The free-streaming stable fixed point is reached as $t\to \infty$, whereas the unstable fixed point is reached when the limit $t\to 0$ is considered\footnote{%
    The unstable fixed point acts as a \textit{stable fixed point} under backward evolution \cite{Chattopadhyay:2021ive, Jaiswal:2021uvv}}. 
For Kasner I, $\chi=-1$, and one can easily see that the stable and unstable fixed points correspond to $\xi = \infty$ and $\xi=-1$, respectively. Similarly, for Kasner II, $\chi=1$, and the stable and unstable fixed points correspond to $\xi = -1$ and $\xi=\infty$ respectively. We define the attractor as the solution which originates from the stable fixed points.

\begin{figure}[t!]
  \centering
  \includegraphics[width=\linewidth]{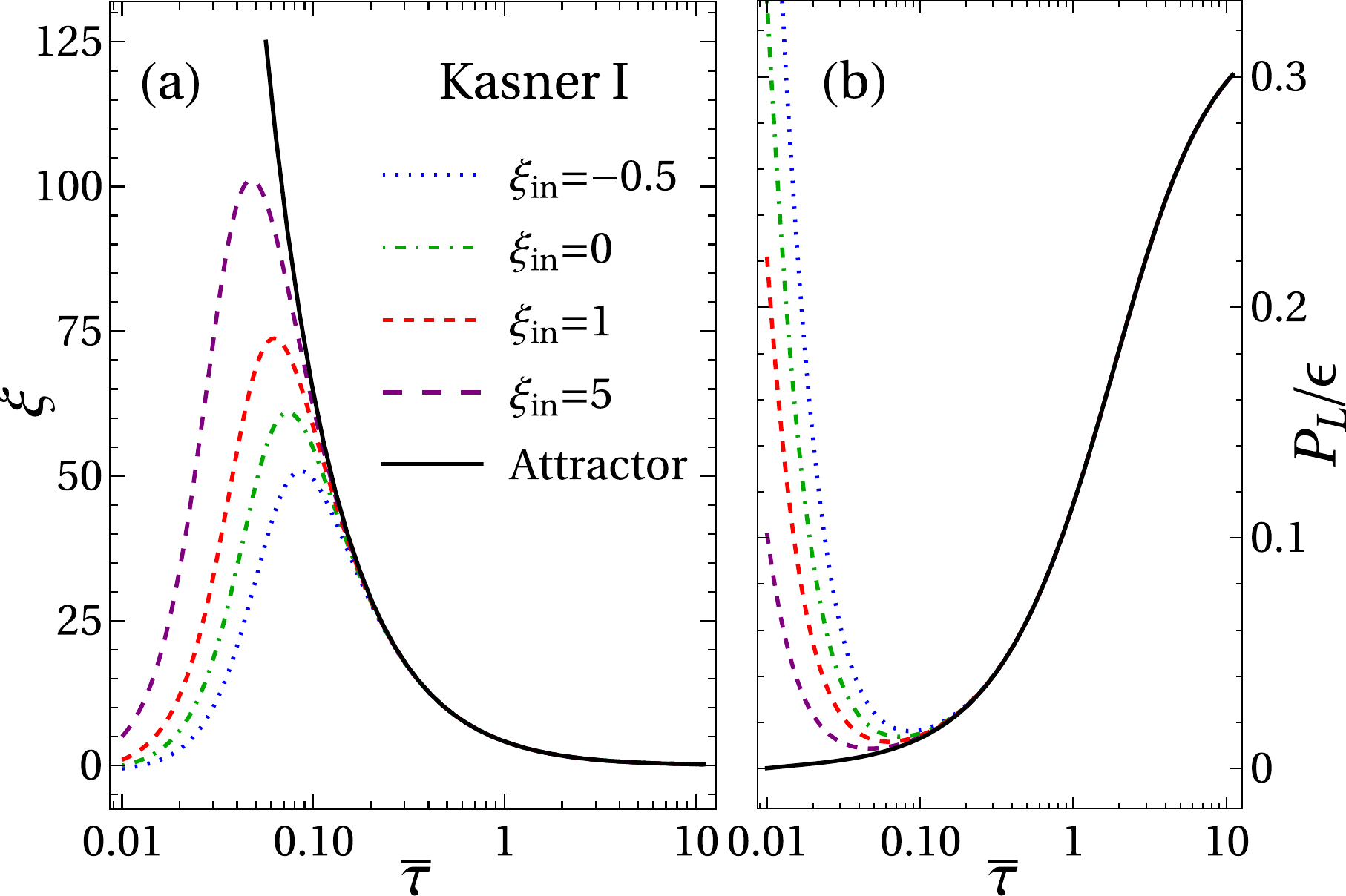}
  \vspace{-.6cm}
  \caption{Evolution of (a) anisotropy parameter $\xi$ and (b) $P_L/\epsilon$ with scaled time. Various initializations of $\xi_{\rm in}$ at $\bar\tau=10^{-3}$ are seen to rapidly approach the attractor solution (initialized with $\xi \to \infty$) in the regime $\tau\ll\tau_R$ when the effect of collisions can be ignored.}
  \label{fig_kasner1}
  \vspace{-.25cm}
\end{figure}

The evolution of anisotropy parameter $\xi$ and the scaled longitudinal pressure $P_L/\epsilon$ with scaled time $\bar\tau$ is shown in Fig.~\ref{fig_kasner1} for Kasner I case. In Fig.~\ref{fig_kasner1}(a), we see a sharp rise in the anisotropy as the metric expands longitudinally, resulting in the distribution function quickly taking an oblate shape. The rapid power-law decay in $P_L/\epsilon$ is shown in Fig.~\ref{fig_kasner1}(b). It should be noted that a similar far-off-equilibrium attractor also emerges in the transverse pressure since for conformal systems the pressures are linearly related, $T^\mu_\mu = \epsilon-2P_T-P_L=0$. However, as emphasized in Refs.~\cite{Chattopadhyay:2021ive, Jaiswal:2021uvv, Jaiswal:2022udf}, the far-off-equilibrium attractor is driven by the rapid longitudinal expansion and hence manifests in $P_L$.

\begin{figure}[t!]
  \centering
  \includegraphics[width=\linewidth]{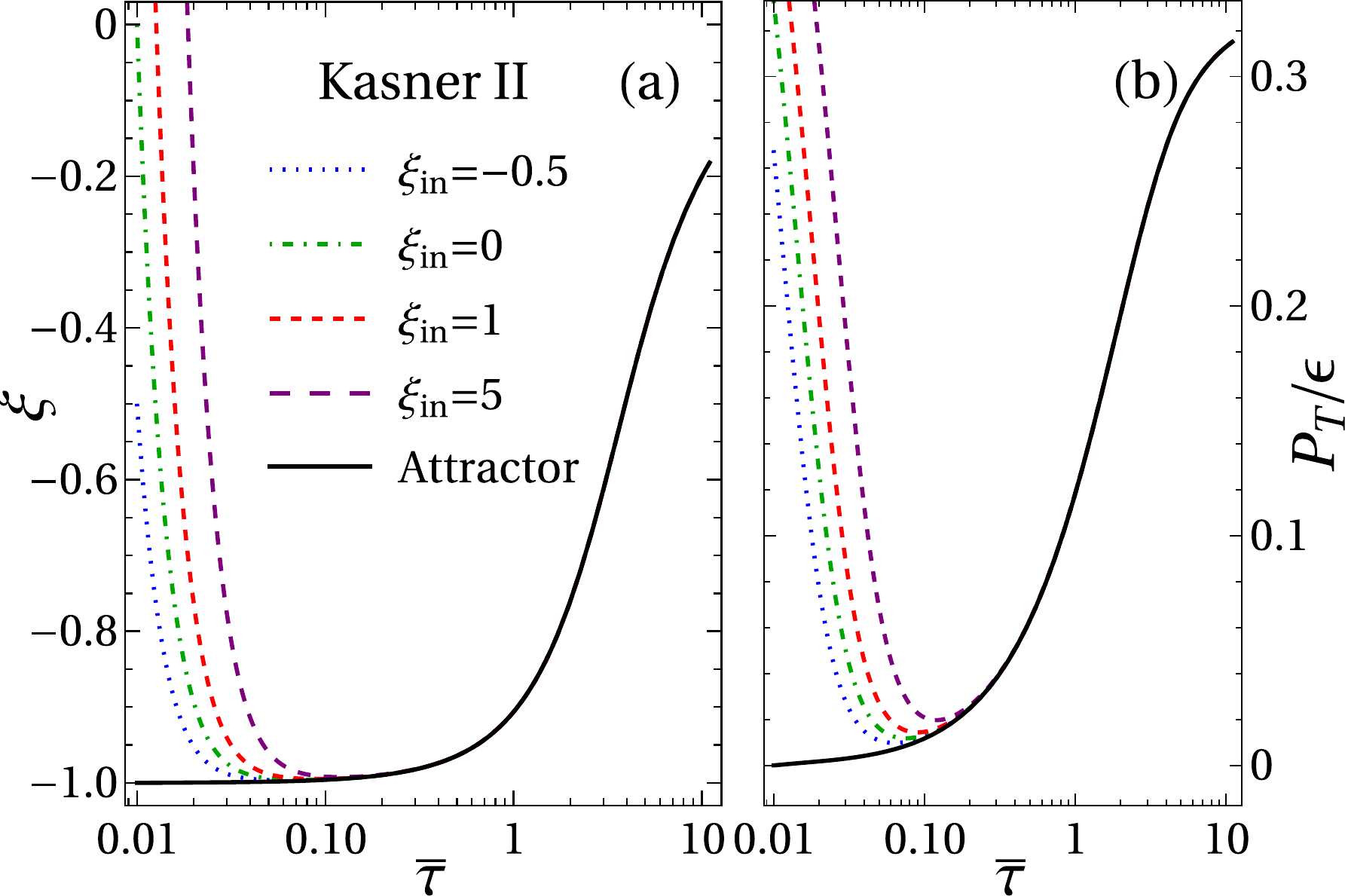}
  \vspace{-.6cm}
  \caption{Evolution of (a) anisotropy parameter $\xi$ and (b) scaled transverse pressure $P_T/\epsilon$ with scaled time. Various initializations of $\xi_{\rm in}$ at $\bar\tau=10^{-3}$ rapidly decay on the attractor solution (initialized with $\xi \to -1$). Far-off-equilibrium universality can be seen in $P_T/\epsilon$ in panel (b).}
  \label{fig_kasner2}
  \vspace{-.2cm}
\end{figure}

In Fig.~\ref{fig_kasner2}, we show the evolution of the anisotropy parameter $\xi$ and $P_T/\epsilon$ with scaled time $\bar\tau$ for Kasner II case. In this case, the rapid \textit{transverse} expansion drives the distribution function to quickly take a prolate shape. This can be seen from Fig.~\ref{fig_kasner2}(a); different initializations quickly decay onto the far-off-equilibrium attractor. As collisions start to become relevant (around $\tau \sim \tau_R$) and the system equilibrates, the anisotropy evolves towards $\xi=0$. The attractor also manifests in the hydrodynamic quantity $P_T/\epsilon$ in Fig.~\ref{fig_kasner2}(b).

From the two Kasner cases, it is apparent that the initial dynamics of the systems are governed by free-streaming in the regime $t \ll \tau_R$; the solutions quickly evolve towards the free-streaming stable fixed point, i.e., towards longitudinal (transverse) free-streaming for Kasner I (Kasner II) case. In Kasner I case, the z-directional expansion causes the distribution to take an oblate shape, while the transverse expansion in Kasner II case makes the distribution have a prolate shape. At time $t \sim \tau_R$, the effect of collisions becomes relevant and drives the system towards local isotropization \cite{Blaizot:2019scw, Jaiswal:2022udf}. Before collisions take over, the degree of anisotropy in the distribution function seems to be governed by the initial time and the rate of expansion. In the next section, we check this hypothesis by considering expansion geometries beyond the Kasner cases, which gives us the freedom to study the dependence of the anisotropic evolution on the expansion coefficients.

\vspace{-.3cm}
\section{Expansion geometry and early-time attractor}
\vspace{-.2cm}

In this Section, we analyze the effects of expansion geometry on the far-off-equilibrium attractor. We consider the axis-symmetric case where $A=B$, with metric scale factors to have the same form as in the Kasner case, i.e., 
\begin{equation}
A = t^a, \qquad B = t^a, \qquad C = t^c,
\end{equation}
but relax the Kasner conditions~\eqref{Kasner_cond} and assume that parameters $(a,c)$ can admit any value. In this case, Eq.~\eqref{first_moment} reduces to
\begin{equation}\label{first_moment_gen}
\frac{\mathcal{R}^{\prime}}{\mathcal{R}} \frac{\partial \xi}{\partial t} + \frac{4}{\Lambda}\frac{\partial \Lambda}{\partial t} = - \frac{(2a+c)}{t} \left(1+\frac{1}{3} \frac{\mathcal{R}_L}{\mathcal{R}} \right).
\end{equation}
Similarly, Eq.~\eqref{second_moment} reduces to,
\begin{equation}\label{second_moment_gen}
\frac{1}{1+\xi} \frac{\partial \xi}{\partial t} + \frac{2(a-c)}{t} + \frac{\mathcal{R}^{5/4}}{\tau_R} \xi \sqrt{1+\xi}=0.
\end{equation}
The above two equations can be combined to obtain an evolution equation for $\xi$ analogous to Eq.~\eqref{eq:attractor} as
\begin{align}\label{eq:gen}
\frac{1}{1+\xi}\!\left[\!1-\frac{2a+c}{12}\!\left(\!\frac{3\mathcal{R}+\mathcal{R}_L}{\mathcal{R}} \!\right)\! \right]\! \frac{\partial \xi}{\partial \bar{\tau}} + \frac{2(a-c)}{\bar{\tau}}
+ \mathcal{R}^{5/4}\xi \sqrt{1+\xi} =0.
\end{align}

\begin{figure}[t!]
  \centering
  \includegraphics[width=\linewidth]{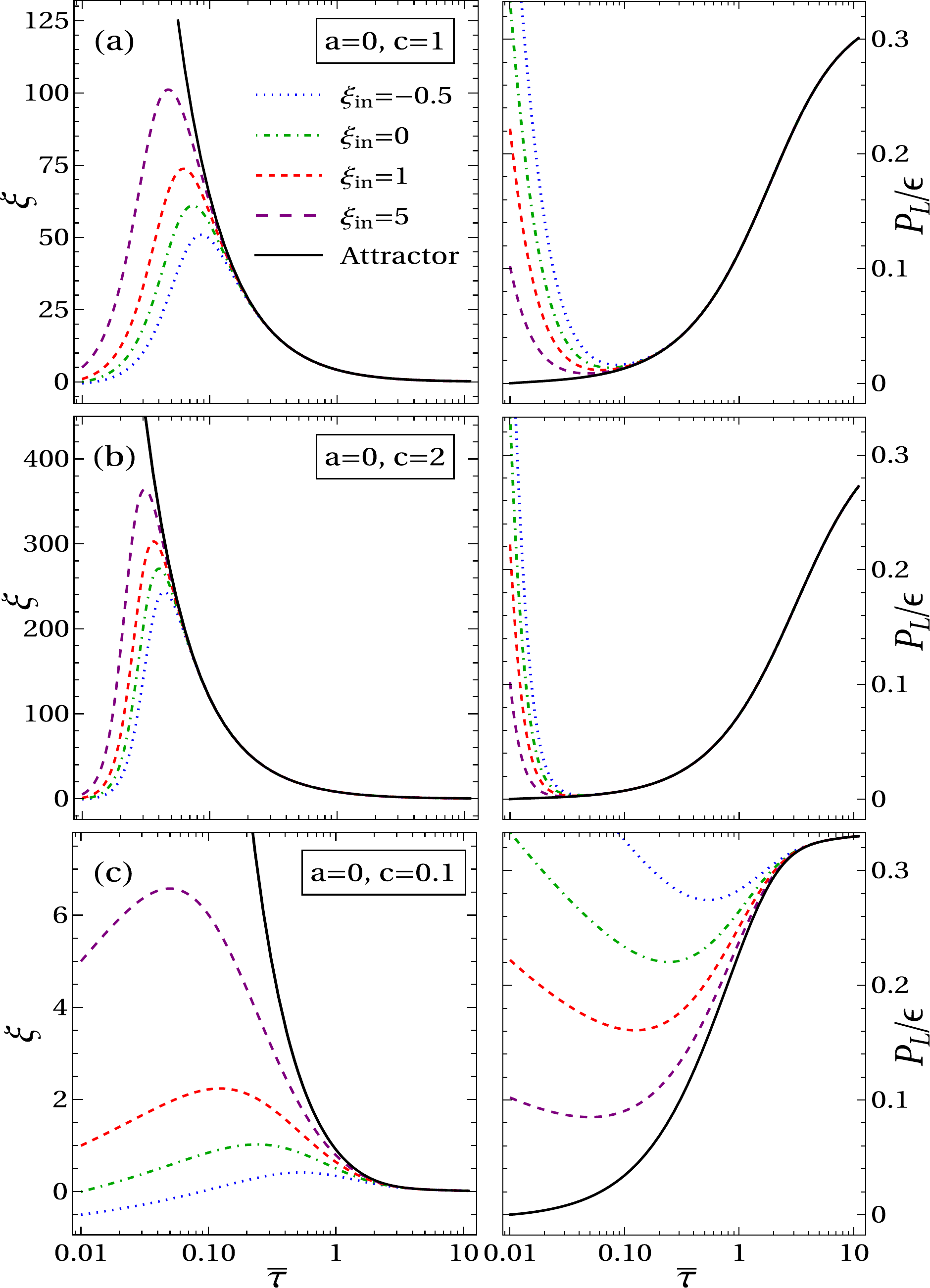}
  \vspace{-.6cm}
  \caption{Evolution of (a) anisotropy parameter $\xi$ and (b) $P_L/\epsilon$ with scaled time for $a=0$ and three different values of the parameter $c$. The effect of the expansion rate is visible: a slower expansion rate delays the approach to the far-off-equilibrium attractor.}
  \label{fig_varc}
    \vspace{-.2cm}
\end{figure}

In absence of collisions, the free-streaming evolution of $\xi(t)$ can be obtained from Eq.~\eqref{second_moment_gen} by taking the limit $\tau_R\to \infty$ as
\begin{equation}\label{decay}
    \xi_{\rm FS}(t) = -1 + \left(\frac{t_{\rm in}}{t}\right)^{{2(a-c)}} (1+\xi_{\rm in}).
\end{equation}
It can be easily seen that the free-streaming evolution of $\xi$ is governed by a power law dependence. The stable fixed point can be obtained by considering the solutions at $t\to \infty$, and is obtained to be $\xi=\infty$ and $\xi =-1$ corresponding to negative and positive values of the exponent $(a-c)$, respectively. The unstable fixed point acts as a stable fixed point under backward evolution of $\xi$ starting at an initial time $t_{\rm in}$, and is obtained to be $\xi=\infty$ and $\xi =-1$ corresponding to positive and negative values of $(a-c)$, respectively. Note that for isotropic metric expansion, i.e., $a=c$, the solution of Eq.~\eqref{decay} is just $\xi_{\rm FS}(t) = \xi_{\rm in}$, which is expected as the anisotropy is not affected by the isotropic expansion geometry. We shall now study the approach of different initialization of $\xi$ to the far-off-equilibrium attractor in presence of collisions.

\subsection{Effect of varying longitudinal expansion rate}

\begin{figure}[t!]
  \centering
  \includegraphics[width=\linewidth]{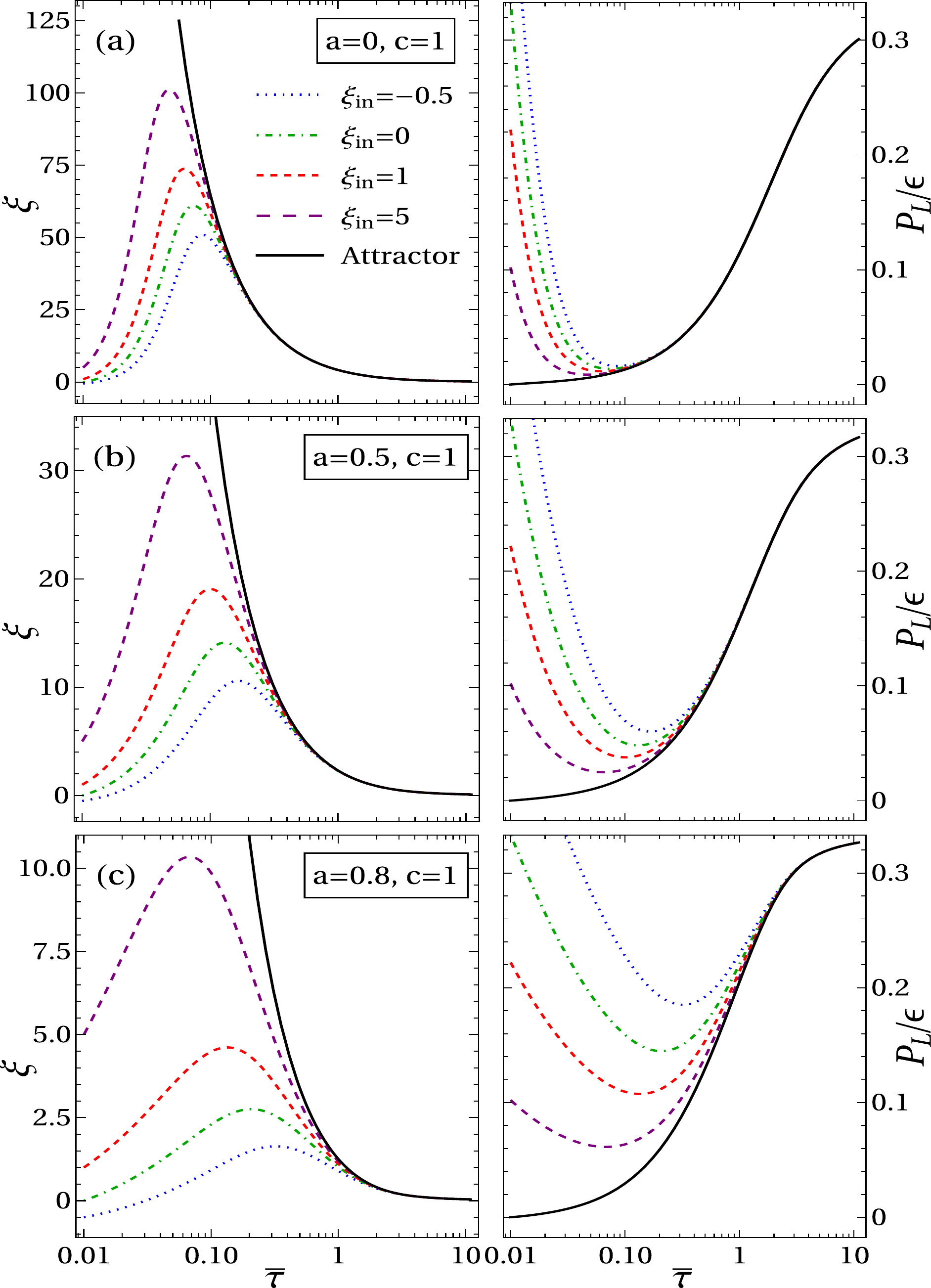}
  \vspace{-.6cm}
  \caption{Evolution of (a) anisotropy parameter $\xi$ and (b) $P_L/\epsilon$ with scaled time for $c=1$ and three different values of the parameter $a$. The effect of the expansion rate is visible: faster transverse expansion delays the approach to the far-off-equilibrium attractor. }
  \label{fig_vara}
\vspace{-.2cm}
\end{figure}

\begin{figure*}[bht!]
  \centering
  \includegraphics[width=\linewidth]{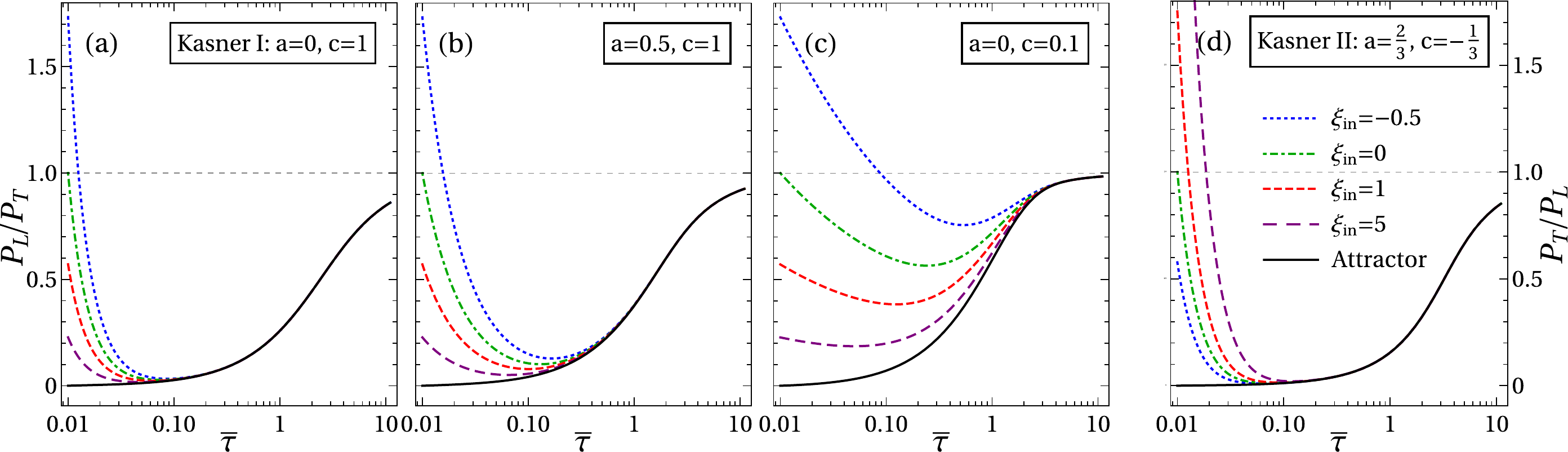}
  \vspace{-.6cm}
  \caption{Evolution of pressure anisotropy with scaled time for different values of the parameters $a$ and $c$. Panels (a), (b), and (c): as the difference in expansion rates reduce, it takes fewer collisions for the system to isotropize. Consequently, $P_L/P_T$ tends to unity faster as $|a-c|$ reduces. The evolution of $P_T/P_L$ is shown in panel (d) for the Kasner II case.}
  \label{fig_6}
  \vspace{-.2cm}
\end{figure*}

As discussed in the preceding Section, the two Kasner cases hint that the existence of far-off-equilibrium relies on rapid expansion geometry. In Fig.~\ref{fig_varc}, we set the metric scale factor $a=0$, which implies that there is no expansion in the transverse direction. The stable fixed point in this case corresponds to $\xi=\infty$ as all free-streaming solutions take this value asymptotically. Panel (a) shows the Kasner I case, or equivalently, Bjorken expansion. As already discussed, the far-off-equilibrium universality is seen for the anisotropy parameter $\xi$, and the early-time attractor is seen for the scaled longitudinal pressure. The expansion rate is increased by setting $c=2$ in Fig.~\ref{fig_varc}(b). A faster convergence compared to panel (a) is observed for the anisotropy, as the power law decay depends on the value of $(a-c)$ (see Eq.~\eqref{decay}). This feature is reflected in $P_L/\epsilon$ in the right panel of Fig.~\ref{fig_varc}(b). However, if the expansion rate is decreased, no far-off-equilibrium attractor is seen in the anisotropy parameter as well as $P_L/\epsilon$, as can be seen from Fig.~\ref{fig_varc}(c) where we have set $c=0.1$. This demonstrates that the far-off-equilibrium attractor is indeed driven by the rapid expansion geometry as suggested in \cite{Chattopadhyay:2021ive, Jaiswal:2021uvv, Jaiswal:2022udf}.

\vspace{-.1cm}
\subsection{Effect of varying transverse expansion rate}

Having established that the far-off-equilibrium attractor is a feature of the expansion and manifests in the direction the expansion dominates, we now explore whether such a feature is observed in systems where the transverse expansion is also present. This may be relevant to heavy-ion collisions at lower energies where the transverse expansion becomes sizable%
    \footnote{One should be careful that at lower energies, a conformal kinetic description of the system as well as the $f_{\rm RS}$ ansatz for the distribution function may not be applicable.}. 
In Fig.~\ref{fig_vara}, we fix the longitudinal expansion by fixing the parameter $c=1$, and vary the transverse  expansion by considering three different values of the parameter $a$. As can be seen from panels (a), (b), and (c), increasing the transverse expansion rate delays the approach to the attractor. The presence of early-time attractor thus relies on the \textit{difference of the expansion rates} and not just rapid expansion along a particular direction. As a result, far-off-equilibrium universality is not expected in low energy collisions, where the transverse expansion may become comparable to the longitudinal expansion before the early-time attractor is reached. In high energy collisions, Bjorken expansion ($a=0,\,c=1$) holds for early time when the transverse expansion has not developed and therefore one may expect a far-off-equilibrium universality in this case.

We show the evolution of pressure anisotropy in Fig.~\ref{fig_6}. The difference between the expansion rates in panels (a), (b), and (c) is $1, 0.5,$ and $0.1$, respectively. We observe that the pressure anisotropy, $P_L/P_T$ tends to unity faster as the difference in expansion rates, $|a-c|$, is smaller. 
Also shown in panel (d) is the evolution of $P_T/P_L$ for the Kasner II case. The approach towards transverse free-streaming is seen at early times due to the rapid transverse expansion and longitudinal contraction.

\vspace{-.3cm}
\section{Summary}
\vspace{-.2cm}

We have presented an alternate derivation of anisotropic hydrodynamic equations by considering interacting static fluid in a locally anisotropic background space-time metric given by Bianchi type I metric. We further considered the special case of axis-symmetric Kasner metric and discussed the implications of our results in the context of anisotropic hydrodynamics. We then studied the emergent far-off-equilibrium attractor for the axis-symmetric Kasner case. Further, by relaxing the Kasner conditions on metric parameters, we analyzed the effect of expansion geometries on the far-off-equilibrium attractor. We demonstrated that the approach to the far-off-equilibrium attractor is governed by relative expansion rates along the longitudinal and transverse axes. 

At this juncture, we would like to elaborate on the techniques developed in the present article to study the effects of various hydrodynamic expansion profiles on the early-time attractor. It was demonstrated earlier that a non-interacting static fluid in a locally anisotropic background metric appears to be governed by aHydro equations to an asymptotic observer \cite{Dash:2017uem}. Therefore the anisotropy in the metric is analogous to fluid expansion profile. Inclusion of microscopic interactions in the present article enabled us to study the evolution of an out-of-equilibrium system and its effect on the dynamical attractor. We emphasize that these findings are robust and do not depend on the choice of moment closure for aHydro equations. We have also discussed the relevance of these results in the context of relativistic heavy-ion collisions.

\vspace{-.3cm}
\section*{Acknowledgements}
\vspace{-.2cm}
Authors acknowledge useful discussions with Jean-Paul Blaizot and Ashutosh Dash. A.J. was supported in part by the DST-INSPIRE faculty award under Grant No. DST/INSPIRE/04/2017/000038.

\bibliographystyle{elsarticle-num}
\bibliography{reference}

\end{document}